\documentclass[11pt]{article}
\usepackage{psfig}
\usepackage{amssymb}

\begin{document}
\newcommand{\real}{\mbox{{\bf I}\kern-.15em{\sf R}}}
\newcommand{\ganz}{\mbox{{\bf I}\kern-.15em{\sf Z}}}
\newcommand{\forget}[1]{}
\newcommand{\eq}[1]{\begin{equation} #1 \end{equation}}
\newcommand{\ma}[1]{$ #1 $}

\newcommand{\beq}{\begin{displaymath}}
\newcommand{\eeq}{\end{displaymath}}
\newcommand{\beqn}{\begin{equation}}
\newcommand{\eeqn}{\end{equation}}
\newcommand{\beqa}{\begin{eqnarray}}
\newcommand{\eeqa}{\end{eqnarray}}

\begin{flushright}
ITFA 99-01\\
January 1999
\end{flushright}

\begin{center}
\vspace{24pt}

{\Large \bf Hypercubic Random Surfaces with Extrinsic Curvature} 
\vspace{24pt}

{\large \sl S.~Bilke}\\
\vspace{10pt}
   Institute  for Theoretical Physics, University of Amsterdam \\
    Valckenierstraat 65, 1018 XE Amsterdam, The Netherlands\\
\vspace{10pt}

\begin{abstract} 
We analyze a model of hypercubic random surfaces with an extrinsic
curvature term in the action. We find a first order phase transition 
at finite coupling separating a branched polymer from a stable flat phase. 
\end{abstract}
\end{center}
\vspace{15pt}

\section{Introduction}
Random surfaces have attracted a lot of interest in different branches
of physics in the recent years. The thermodynamics  
of surfaces embedded in  ``real'' three-dimensional Euclidean space is
interesting in 	itself as it may describe some properties of real membranes
appearing in nature.   In high energy physics random surfaces play a role for 
example  as the world-sheet of strings or as the space-time of two-dimensional
quantum gravity. In this paper we investigate the hyper-cubic random surface
model with an extrinsic curvature term in the action. 
Although  originally motivated by bosonic string theory it may well be 
interpreted as a fluid phantom membrane with bending rigidity. 
The word phantom refers to the fact that the surface considered here can,
differently form real-world surfaces, self-intersect. 

The hypercubic random surface model was originally proposed by Weingarten 
\cite{dw80} as a non-perturbative regularization of the world-sheet of 
bosonic strings. It was, however, soon realized that the dominating 
geometries in this model have the essentially one dimensional structure of 
branched polymers. It was shown \cite{dfj84} that for spherical surfaces the 
string susceptibility exponent $\gamma $ is equal to $1/2$, 
the generic value for branched polymers. In \cite{dj86} the model was extended
to include an extrinsic curvature term and it was shown, under some 
assumptions, such a term in the action does not change the model's critical
behavior for any finite coupling. It therefore came as a surprise when 
numerical evidence for a non-trivial 
behavior with $\gamma = 1/4$ was observed \cite{bb86} when an additional 
local constraint on the allowed configurations was introduced,  a term
forbidding self-bending surfaces. The result $\gamma = 1/4$ was especially 
interesting because this value fits nicely into a series  
$\gamma = 1 /n, n=2,\cdots$ of positive string
susceptibility exponents discussed in \cite{du94}. Hence this result led to 
some speculations \cite{a94}. However, using improved numerical techniques
and larger lattice-sizes it was shown  \cite{myself_hyp} that the true 
large volume behavior is masked by strong finite size effects but 
nonetheless is well described by branched polymers with $\gamma = 1/2$. 

In this paper we generalize the hyper-cubic random surface model by adding 
a term coupled to the external curvature to the action. 
Introducing such a term will certainly change the dominating geometry, at 
least in the infinite coupling limit. In this limit the dominating geometry is
flat with external Haussdorff dimension\footnote{for a definition see eq. 
\ref{how_hau}} $d_h = 2$ different from $d_h = 4$ for 
branched polymers. We will demonstrate numerically that the transition
actually happens at {\em finite} coupling. This means we observe for a 
fluid membrane a crumpling transition separating the branched polymer from 
a stable flat phase. 

On the first glance, the announced  transition at
finite coupling seems to be ruled out by \cite{dfj84}, which states that
``the critical exponents take their mean field value if the 
susceptibility of the model and a coarse grained version of it both diverge''.
In other words under the assumptions
 \eq{\gamma > 0 \quad \mbox{and }  \bar{\gamma} > 0\label{assumptions}} 
the model is always in the branched polymer phase for finite couplings. 
This result was obtained with a renormalization group 
argument where the branched polymer was decomposed into "blobs", components 
which can not be cut into two parts along a loop of length two. 
A key step in the derivation of this result in \cite{dj86} is the formula
\eq{\chi (\mu ) = \frac{\bar{\chi}(\bar{\mu})}{1-\bar{\chi}(\bar{\mu})},
\label{durhuus_grund}} 
which relates the susceptibility $\chi(\mu)$ of the original model to the 
susceptibility $\bar{\chi}(\bar{\mu})$ of the decomposed blobs with
renormalized coupling $\bar{\mu}$. Obviously $\bar{\chi} = 1$ if the 
susceptibility $\chi $ diverges. Under the assumptions stated above, 
i.e. $\bar{\gamma} > 0$ for the blobs  one concludes that 
the blobs are not critical $\bar{\mu} > \bar{\mu}_c$ and $\bar {\chi}$ is 
analytic at this point. With a Taylor expansion one gets the self consistency 
relation $ \gamma = 1 - \gamma $ which is solved by $\gamma = \frac{1}{2}$,
the generic value for branched polymers.  However, Durhuus emphasized 
\cite{du94} that the condition $\bar{\chi} = 1$ alone does not imply that 
the blobs are  non-critical. If $\bar{\gamma} < 0$ the susceptibility 
$\bar{\chi}$ does not diverge and $\bar{\chi} = 1$ can be satisfied {\em at}
the critical point $\bar{\mu} = \bar{\mu}_c$ of the coarse grained system. 
At such a point the entropy is still dominated by the branching of the 
surface but the branches themselves are critical, which in effect changes 
the exponent $\gamma$ for the whole system.  If one assumes that 
$\bar{\gamma}$ takes the KPZ-values \cite{kpz} one can derive 
from  (\ref{durhuus_grund}) a series of positive 
$\gamma = 1/n, n=2,\cdots$ \cite{du94}. 

The derivation of this series is rather formal. It does not provide a
description how to construct a system which has this property. We want to
check numerically if the hyper-cubic random surface model exhibits non trivial
behavior at an eventual crumpling transition. In other words we check
if the assumptions (\ref{assumptions}) are fulfilled in the whole
range of couplings $\epsilon$\footnote{see eq. (\ref{ext_curv})}. 
Non-trivial behavior can be expected only if 
these assumptions are not satisfied at a possible phase transition induced
by the external curvature coupling.
A special focus  will therefore be given  to a possible transition 
point.

\section{The model}
A hyper-cubic surface is an orientable surface 
embedded in $Z^D$, obtained by gluing pairwise together plaquettes $P$ 
along their links until no free edges are left. A plaquette $P$ is a 
unit-square which occupies one of the unit-squares 
$S_{z^{\mu}}, \mu = 1 \cdots D$  in the embedding lattice. 
In the following we  use the word square to refer to the embedding lattice 
and the words link or plaquette to describe the internal
connectivity of the surface. A link is shared by exactly two plaquettes, which
are glued together along each one of their edges. Note that a link or a 
square in the embedding lattice can be occupied  more than once which means
these are non self-avoiding, {\em i.e.} phantom, surfaces. 

The partition function is
\eq{
{\cal Z}(\beta)= \sum _{E \in \cal S} e^{- S} 
        = \sum _A e^{-S} {\cal N}(A), 
\label{hyper_Z}
}
where the sum runs over an ensemble $\cal S$ of hyper-cubic surfaces 
with fixed (spherical) topology. The number ${\cal N}(A)$
of surfaces  with a given area $A$ is expected to grow like 
\eq{ 
{\cal N}(A) \approx e^{\mu _{c} A} A ^{\gamma - 3} 
\label{hyper_largeN}
}
in the large volume limit. The coefficient $\gamma $ is the entropy exponent. 
To balance the exponential growth of the number of surfaces with the volume,
a term $\mu A$ linear in $A$  has to appear in the action. The partition 
function is well-defined only if $\mu \le \mu_c$. For the extrinsic 
curvature  we note that there are three possible ways 
to embed two neighboring plaquettes $p_1, p_2$
in a hyper-cubic lattice $Z^D$. The external angle $\Theta _l$, which we
assign to link $l$, can therefore take three possible values:
\begin{itemize}
\item[0: \hspace{3mm}] 
             $p_1$ and $p_2$ occupy two neighboring squares in the same 
             coordinate plane,
\item[$\frac{\pi}{2}$: \hspace{3mm}]  
             $p_1$ and $p_2$ occupy two neighbor squares in 
             different coordinate planes,
\item[$\pi$: \hspace{3mm}] 
             $p_1$ and $p_2$ occupy the same square. We call such a
             configuration self-bending. 
\end{itemize}
We use the potential
\eq{
S_{ext}[E](\epsilon)  
      =  \mu A + 
        \frac{\epsilon}{2} \sum _{l \in E} \left ( 1 - \cos \Theta _l \right ) 
      =  \mu A +
 \epsilon \sum _{l \in E} \left ( \delta _{\pi, \Theta _l} + 
                      \frac{1}{2} \delta _{\pi / 2 , \Theta _l} \right )
\label{ext_curv}.}

The physical questions we want to address are most conveniently analyzed 
using the canonical ensemble. However, the algorithm used for this work
\cite{myself_phd} requires moderate volume fluctuations for ergodicity. 
Therefore we simulate the quasi-canonical ensemble:
\eq{
Z(\overline{A}, \epsilon) = 
   \sum _{E \in {\cal S}_{\overline A}} e^{-S[E](\overline{A},\epsilon)}
               = \sum _{E \in {\cal S}} \delta_{A, \overline{A}} 
                 e^{-S[E](\overline{A},\epsilon) - 
                     \frac{\gamma}{2}\left ( \overline{A} - A \right )^2} 
\label{hyper_z}
}  
In the first expression, defining the canonical ensemble, the sum runs over all
surfaces with area $\overline{A}$. In the actual simulation surfaces with 
areas different from $\overline{A}$ are generated, where the area-fluctuations
are confined by the additional Gaussian potential. The information about the
canonical ensemble is extracted by measuring only if $A = \overline{A}$.

\begin{figure}
\hspace{2cm}
\psfig{file=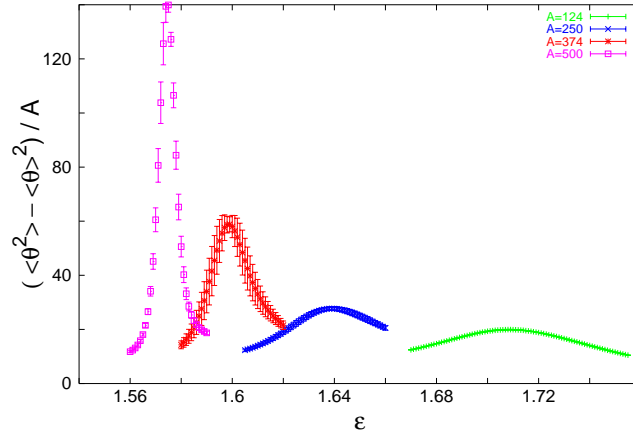,angle=270,height=8.5cm,rheight=5.5cm}
\caption{The susceptibility $\chi _{ext}$ in $D=3$ dimensions}
\label{chi_ext3}
\end{figure}

\subsection{Numerical results}
\begin{figure}
\hspace{2cm}
\psfig{file=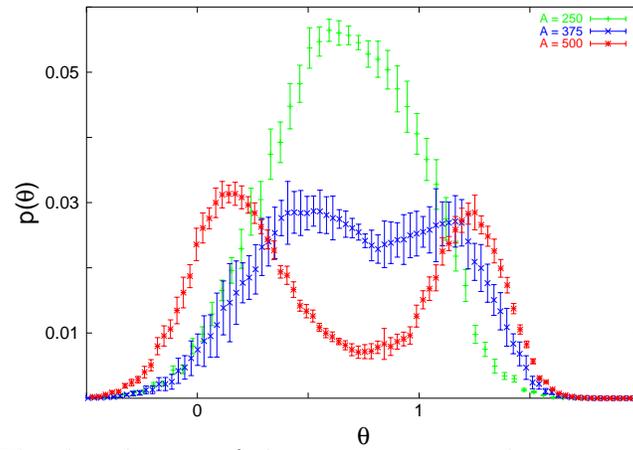,angle=270,height=8.5cm,rheight=5.5cm}
\caption{The distribution of the average external curvature $\bar{\Theta}$ in
$D=3$ dimensions obtained with re-weighting. The non zero probability for 
unphysical negative values $\bar{\Theta} < 0$ is an artifact of the 
extrapolation to this region in the re-weighting process.
}
\label{eden}
\end{figure}
To localize  possible transitions we measure the energy fluctuation 
of the external curvature field
\eq{\chi_{ext} = \frac{1}{A} \frac{\partial ^2}{\partial ^2 \epsilon} 
           \log Z(A,\epsilon) = \frac{<\Theta^2> - <\Theta>^2}{A}} 
and search for maxima in this observable. The average external curvature is
proportional to the first derivative of the free energy~:
\eq{ <\Theta> =  \frac{\pi}{2 A} \frac{\partial }{\partial \epsilon} 
                     \log Z(A, \epsilon) 
= \frac{1}{2 A} \sum _S \sum _l \Theta _l e^{-S[E]}.
}

We simulated in three and four  embedding dimensions and scanned the 
coupling range $0 \le \epsilon \le 3$ for maxima of $\chi _{ext}$. 
In both cases we found a single peak. We used re-weighting methods 
\cite{FS88} to extract the shape of the peaks shown in figure \ref{chi_ext3} 
from four independent
measurements per volume at $\epsilon \approx \epsilon _{pc}$ in $d=3$.
 
The peaks grow quickly,  in fact a bit faster than
linear, with the volume. This suggests a first order phase
transition, which is confirmed by a look at the distribution of 
external curvature $\bar{\Theta}$, where the bar indicates the average taken 
over a given lattice. In figure \ref{eden} we show  this distribution
for $D=3$ and the coupling $\epsilon \approx \epsilon_{pc}$ close
to the pseudo-critical coupling. One observes a clear signal of a first order 
phase transition, namely  two maxima separated by a minimum which becomes
deeper as  the size of the surfaces is increased. 
This clearly indicates two separate phases. 
In one phase the average external curvature  is close to zero, the typical
surface is flat. 

\begin{figure}
\hspace{3cm}
\psfig{file=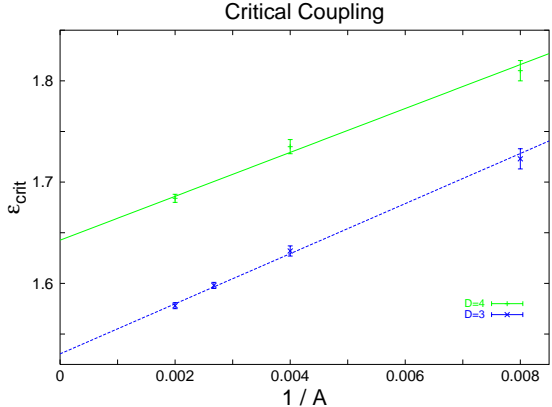,angle=270,height=7.5cm,rheight=5cm}

\caption{The pseudo-critical coupling $\epsilon _{crit}(1/A)$. }
\label{eta_crit}
\end{figure}
\noindent
To estimate the critical coupling we use that for a first order phase 
transition one can expect 
\eq{\epsilon _{pc}(A) = \epsilon_c + \frac{c}{A} + O(\frac{1}{A^2}) 
\label{SKL}}
for the scaling of the pseudo-critical couplings $\epsilon _{pc}$. 
In figure \ref{eta_crit} we show the numerical estimates  for  
$\epsilon _{pc}$. Note that the pseudo critical coupling decreases with 
the volume which indicates  that $\epsilon _c$ is finite in the 
thermodynamic  limit. 
With a linear fit to equation (\ref{SKL}) we find for the critical coupling

\begin{displaymath}
\begin{array}{cc}
  D = 3 & \epsilon _c = 1.530(3) \\
  D = 4 & \epsilon _c = 1.643(4) 
\end{array}.
\end{displaymath}

\forget{
\begin{figure}
\psfig{file=gy.ps,angle=270,height=7cm,rheight=5cm}
\caption{The scaling of the radius of gyration. }
\label{hausdorff}
\end{figure}
}
\begin{figure}
\hspace{2cm}
\psfig{file=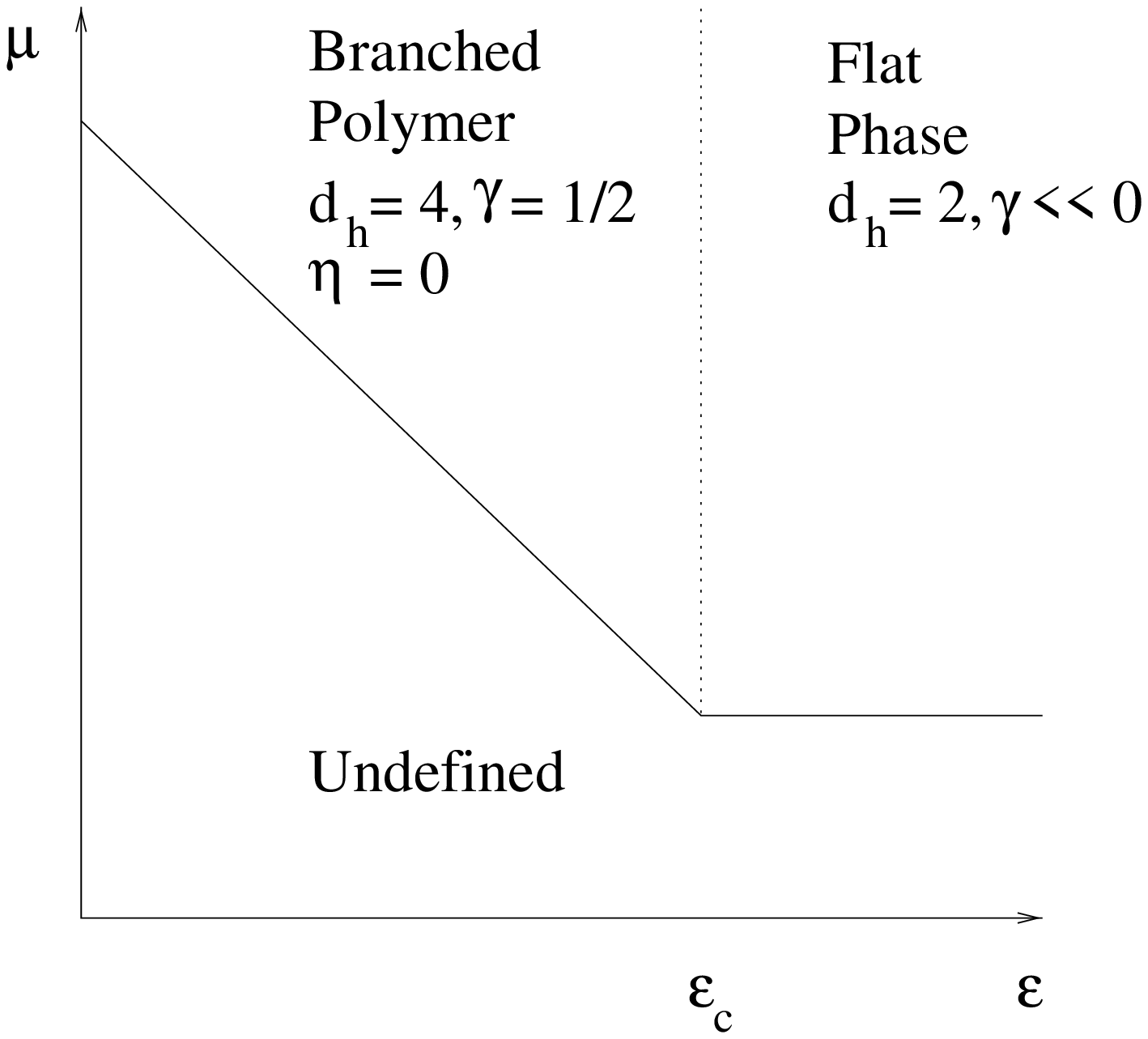,angle=270,height=7cm,rheight=7cm,rwidth=5cm}

\caption{The schematic phase diagram for the Hyper-cubic random surface 
model with
extrinsic curvature. One finds qualitatively the same behavior in $D=3$ and 
$D=4$ embedding dimensions.} 
\label{hyper_phase}
\end{figure}

To demonstrate the change in the geometrical behavior we consider  the
radius of gyration
\eq{G^2(A, \epsilon) = < \frac{1}{A} 
          \sum _p (S_{\mu}(p)  - S_{\mu}(p_0))^2>, }
which is the average squared distance of the plaquettes to some reference
plaquette $p_0$.  The $S_{\mu}$ are the coordinates of the plquette in the
embedding space. In the numerical simulations we do a stochastic average over
the $p_0$ by choosing 10 \% of the plaquettes at random as $p_0$ 
per measurement. One can use the radius of gyration  to define the 
external Haussdorff dimension $d_h$
\eq
{G^2(A, \epsilon) \approx  A^{2 / d_h}. \label{how_hau} }
In a sense the Haussdorff dimension is the largest possible  dimension for 
the embedding space  which can still be completely filled by the surface. 
We have measured numerically the radius of gyration for $\epsilon = 0$ 
and $\epsilon = 2$ for volumes $A$ in the range $200$ to $5000$. With a fit
to (\ref{how_hau}) we extract the Haussdorff dimension:
\begin{displaymath}
\begin{array}{cc}
\epsilon = 0.  & d_h = 4.07(3) \\
\epsilon = 2.  & d_h = 2.00(3).   
\end{array}
\end{displaymath}
The value found for $\epsilon = 0$ is two standard deviations away from the
expected result \cite{bb86} but this should presumably be attributed to 
finite size effects. 

The results are summarized in the phase diagram figure \ref{hyper_phase}.
The model is defined only for $\mu > \mu _c(\epsilon )$ which defines 
the critical line depicted by the solid line in the phase diagram. 
Above this line we find two phases. 
For $\epsilon < \epsilon _{crit}$ the system is in the branched polymer phase 
with $d_h \approx 4$. If we assume
hyperscaling \eq{ \nu = d_H^{-1}, \quad \gamma = \nu (2 - \eta)}
the anomalous scaling dimension is $\eta = 0$ as expected for a branched
polymer. The strong coupling phase is flat, which is reflected by the
observation $d_h = 2$. In this phase basically no baby universes  are present. 
For example, on a surface of size $A=500$ 
we found at $\epsilon = 2$ on average only $0.06$  baby universes  of 
size $B=5$. The largest one found in this phase is of size 
$33$ and appeared only once in $200 000$ measurements. For comparison in the 
branched polymer phase at $\epsilon = 0$ we found in average $10.6$ baby
universes  of size $5$ on the surface, the probability for the largest 
possible one with $B=249$ is about $0.02$.  Therefore one cannot use the 
baby universe distribution to measure $\gamma$ in the flat phase. However the 
fact  that there effectively are no baby surfaces means that $\gamma $ is 
large negative. Therefore the existence of this phase is not 
in contradiction with \cite{dj86} because the assumptions (\ref{assumptions}) 
are not met. 

\section{Summary}
We have analyzed the model of hypercubic random surfaces with an 
extrinsic curvature term in the action. We observed a first order 
phase transition, which separates a branched polymer phase from a flat phase.
We have shown that the critical coupling is finite in the thermodynamic
limit, i.e. the flat phase survives in the large volume limit and we observe
true long-range order. This may seem a bit surprising
first because long range correlations in two dimensional systems are rather 
unusual. In a comparable model of triangulated fluid membranes the existence
of a stable, long range ordered phase had somewhat been disputed 
\cite{disputed}.

Our hope to find a non-trivial, positive $\gamma$ for the geometry at 
the transition of the external-curvature ``field'' was not satisfied. 
Although we find a phase transition, we do not have the situation where
the exponent $\overline{\gamma}$ of the individual branches is negative
while the exponent for the overall geometry is positive but 
different from the generic branched polymer value $\gamma = \frac{1}{2}$.
Instead we find $\gamma $ large negative, which is also the reason why 
the transition is not in contradiction with the statement ``no transition at 
finite coupling'' for this model given in \cite{dj86}, because this 
argument relies on the assumption, that $\gamma $ is positive.

\end{document}